# SELF-CONSISTENT METHODS IN NUCLEAR STRUCTURE PHYSICS

JACEK DOBACZEWSKI

*Institute of Theoretical Physics, Warsaw University*
*ul. Hoża 69, PL-00681 Warsaw, Poland*
*E-mail: dobaczew@fuw.edu.pl*

*Joint Institute for Heavy Ion Research, Oak Ridge National Laboratory,*
*P.O. Box 2008, Oak Ridge, TN 37831, U.S.A.*

*Department of Physics, University of Tennessee,*
*Knoxville, TN 37996, U.S.A.*

We present a very brief description of the Hartree-Fock method in nuclear structure physics, discuss the numerical methods used to solve the self-consistent equations, and analyze the precision and convergence properties of solutions. As an application we present results pertaining to quadrupole moments and single-particle quadrupole polarizations in superdeformed nuclei with $A{\sim}60$.

## 1 Introduction

Self-consistent methods have been used in the low-energy nuclear structure studies over many years, and represent a mature field with numerous successful applications.[1,2,3,4,5] A number of computer codes solving the nuclear Hartree-Fock (HF) problem have already been developed. Two types of effective nucleon-nucleon interactions have been mainly employed. Starting with the work of Vautherin and Brink[6] many authors have applied the nuclear HF theory with the Skyrme effective interaction, while the work of Gogny[7,8] initiated numerous studies with the force which carries his name.

The methods employed to solve the HF equations depend mainly on the effective force used and on the assumed symmetries of the many-body wave functions. For the solutions which allow at least triaxial deformations, two different methods have been applied for the two above mentioned effective interactions. The first one, used in conjucture with the Skyrme interaction, is formulated in the spatial coordinates and makes use of the finite-difference,[9] or Fourier,[10] or spline-collocation [11,12] methods to approximate differential operators. The solution is then obtained by using the imaginary time evolution operator.[13]

The second one, used for the finite-range Gogny interaction, employs a truncated harmonic oscillator (HO) basis [14,15] and solves the problem either by an iterative diagonalization of the mean-field Hamiltonian, or by the gradient [16] or the conjugate gradient[17] methods. Recently, the method which incorporates the advantages of both existing approaches, and combines the robustness of the Cartesian HO basis with the simplicity of the Skyrme interaction, has been implemented.[18]

The methods using spatial coordinates have several advantages. First of all, various nuclear shapes can be easily treated on the same footing; the same cubic lattice of points in three spatial dimensions is suitable to accommodate wave functions with, in principle, arbitrary deformations restricted only by a specific symmetrization of the lattice. This allows easy studies of systems for which the deformation is not *a priori* known, or is ill defined because of deformation instabilities or a shape coexistence. Secondly, using spatial coordinates allows studies which address the asymptotic form of nucleonic wave functions at large distances. This is particularly important for a precise description of weakly bound nuclei, where the use of spatial coordinates is a necessity.[19] Third, for the Skyrme zero-range interaction, the mean fields are local (apart from a velocity dependence) and can be easily programmed in the spatial coordinates. Last but not least, the treatment of wave functions on large lattices ($12\times12\times12$ is a typical example) is easily amenable to vectorization or parallelization of the algorithm.

Methods using the HO basis have other advantages. Firstly, the basis provides a natural cut-off for many operators which otherwise are unbound and require particularly delicate treatment in the spatial coordinates. This concerns in particular the multipole moment and the angular momentum operators which are often used as constraining operators. For the corresponding constraints the solutions can become unstable when the non-zero probability amplitudes (wave functions) move towards large distances as it is the case for e.g. weakly bound nucleons. Secondly, much smaller spaces are usually required to describe the nuclear wave functions within a given precision. Typically a basis of about 300 HO wave functions is sufficient for most applications. Third, the iterative diagonalization of the mean field Hamiltonian can be used to find the self-consistent solutions, which provides a rapidly converging algorithm, and, last but not least, scalar or superscalar computers can also be used, because the typical sizes of the information handled is smaller and the performance is less dependent on the use of a vector processor.

Detailed discussion of the stability, convergence, and efficiency of methods using the expansion on the Cartesian HO basis has recently been presented together with the description of the HFODD code.[18] In the present communication we give a brief *résumé* of the methods employed (Sec. 2), present tests of the precision and discuss the CPU times required (Sec. 3), and illustrate possible applications by presenting a few recent results obtained for the $A\sim60$ nuclei (Sec. 4).

## 2 Hartree-Fock method

The eigenequations for the HF single-particle Routhians $h'_\alpha$ are called the HF equations,

$$h'_\alpha \psi_{i,\alpha}(\bm{r}\sigma) = e'_{i,\alpha} \psi_{i,\alpha}(\bm{r}\sigma), \tag{1}$$

where the neutron ($\alpha{=}n$) and proton ($\alpha{=}p$) Routhians read

$$h'_\alpha = -\frac{\hbar^2}{2m}\Delta + \Gamma_\alpha + \delta_{\alpha p} U^{\text{Coul}} + U^{\text{mult}} - \omega_y \hat{J}_y. \tag{2}$$

Detailed expressions for the nuclear mean field operators $\Gamma_\alpha$, Coulomb potential $U^{\text{Coul}}$, and multipole constraint potential $U^{\text{mult}}$ are given in Ref.[18] After solving Eq. (1) one calculates the total energy of the system as a sum of the kinetic, Skyrme, Coulomb, and pairing terms

$$\mathcal{E} = \mathcal{E}^{\text{kin}} + \mathcal{E}^{\text{Skyrme}} + \mathcal{E}^{\text{Coul}} + \mathcal{E}^{\text{pair}}. \tag{3}$$

The same energy can be calculated by using the Routhian eigenvalues $e'_{i,\alpha}$ and occupation probabilities $v^2_{i,\alpha}$ as

$$\tilde{\mathcal{E}} = \tfrac{1}{2}\sum_{i,\alpha} v^2_{i,\alpha} e'_{i,\alpha} + \tfrac{1}{2}\mathcal{E}^{\text{kin}} + \mathcal{E}^{\text{pair}} - \mathcal{E}^{\text{rear}} + \tfrac{1}{3}\mathcal{E}^{\text{Coul}}_{\text{exch}} - \mathcal{E}^{\text{mult}}_{\text{corr}} - \mathcal{E}^{\text{cran}}_{\text{corr}}. \tag{4}$$

Here, the rearrangement energy $\mathcal{E}^{\text{rear}}$ results from the density dependence of the Skyrme interaction. Similarly, the Coulomb exchange energy $\mathcal{E}^{\text{Coul}}_{\text{exch}}$ can be considered as resulting from a zero-order interaction term depending on the density as $\rho_p^{-2/3}$. The remaining terms correct for the fact that Routhians contain constraint terms which are not present in the total energy.

The difference between the energies $\tilde{\mathcal{E}}$ and $\mathcal{E}$ is exactly equal to zero when the densities and fields do not change from one iteration to the next one. Hence their difference

$$\delta\mathcal{E} = \tilde{\mathcal{E}} - \mathcal{E} \tag{5}$$

provides a useful measure of the quality of convergence, and is called the stability of the HF energy.

The HF equations (1) can be solved by expanding the single-particle wave functions $\psi_i(\bm{r}\sigma)$ onto the deformed HO wave functions $\psi_{n_x n_y n_z, s_z}(\bm{r}\sigma)$ in the Cartesian coordinates, i.e.,

$$\psi_i(\bm{r}\sigma) = \sum_{n_x=0}^{N_x} \sum_{n_y=0}^{N_y} \sum_{n_z=0}^{N_z} \sum_{s_z=-\frac{1}{2},\frac{1}{2}} A_i^{n_x n_y n_z, s_z} \psi_{n_x n_y n_z, s_z}(\bm{r}\sigma). \tag{6}$$

Here $N_x$, $N_y$, and $N_z$ are the maximum numbers of the HO quanta corresponding to the three Cartesian directions. However, the sums over $n_x$, $n_y$, and $n_z$ are performed over the grid of points which form a pyramid rather than a cube.

The HO wave functions have the standard form

$$\psi_{n_x n_y n_z, s_z}(\boldsymbol{r}\sigma) = \psi_{n_x}(x)\psi_{n_y}(y)\psi_{n_z}(z)\delta_{s_z\sigma}, \tag{7}$$

where

$$\psi_{n_\mu}(x_\mu) = b_\mu^{\frac{1}{2}} H_{n_\mu}^{(0)}(\xi_\mu) e^{-\frac{1}{2}\xi_\mu^2}, \tag{8}$$

and $\xi_\mu = b_\mu x_\mu$ are dimensionless variables scaled by the oscillator constants

$$b_\mu = \sqrt{m\omega_\mu/\hbar}. \tag{9}$$

Polynomials $H_n^{(0)}(\xi)$ are proportional to the standard Hermite orthogonal polynomials $H_n(\xi)$,[20]

$$H_n^{(0)}(\xi) = \left(\sqrt{\pi} 2^n n!\right)^{-\frac{1}{2}} H_n(\xi). \tag{10}$$

Accuracy of the solution of the HF equations with the wave functions expanded onto the Cartesian HO basis, Eq. (6), depends on the three parameters $\hbar\omega_x$, $\hbar\omega_y$, $\hbar\omega_z$ defining the HO frequencies in three Cartesian directions, and on the number $M$ of the HO states included in the basis. In the code HFODD we use the standard prescription [21,22] to chose the HO states included in the basis, namely, the $M$ states with the lowest HO single-particle energies,

$$\epsilon_{n_x n_y n_z} = \hbar\omega_x(n_x + \tfrac{1}{2}) + \hbar\omega_y(n_y + \tfrac{1}{2}) + \hbar\omega_z(n_z + \tfrac{1}{2}), \tag{11}$$

are selected among those which have $n_x \leq N_0$, $n_y \leq N_0$, and $n_z \leq N_0$, where $N_0$ is the fixed maximum number of HO quanta. It should be noted that in general both $M$ and $N_0$ have to be specified to define the basis. Only for large $N_0$, the basis is defined solely by $M$ and does not depend on $N_0$. In this case, the grid of points $(n_x, n_y, n_z)$ defining the states included in the basis forms a pyramid in three dimensions, with the inclined face delimited by the condition $\epsilon_{n_x n_y n_z} \leq$ const. On the other hand, only for small values of $N_0$ the basis is defined solely by $N_0$ and does not depend on the energy cut-off. In this case the corresponding grid of points $n_x n_y n_z$ forms a cube of the size $N_0$. In all intermediate cases the shape of the basis corresponds to a pyramid with the corners cut off, or to a cube with the corners cut off. Usually $N_0$ is chosen large enough so that all the states allowed by the energy cut-off are included in the basis.

## 3  Precision and convergence

The HF equations (1) can be solved by using standard iterative methods[1] which are based on the self-consistency principle. Namely, the self-consistent solution is attained when the mean fields $\Gamma_\alpha$ appearing in the mean-field Routhian $h'_\alpha$ equal to the mean fields calculated from the occupied single-particle states $\psi_{i,\alpha}(\boldsymbol{r}\sigma)$. Therefore, by iteratively calculating the mean fields and using them to obtain new single-particle states one may eventually reach the self-consistent solution. A convergence of such a procedure is not guaranteed and, in fact, in nuclear applications the method such as formulated above *almost always* diverges. The reason for that is the fact that the consecutive corrections of the mean fields turn out to be too violent and lead to too strong modifications of the states. Since early years of practical implementations of the HF method in nuclear structure physics the simple method to remedy this situation has been devised. Namely, only the fraction $f$ of the mean fields calculated from the single-particle states is used in every iteration, and it is combined with the fraction $1-f$ of the old mean fields.

In practical calculations, the optimal value of $f$ is determined by checking the convergence rate. An example of such an analysis is presented in Fig. 1, where the stability energy (5) is shown for the values of $f$ ranging from 30% to 90%. It can be seen that the rate of convergence increases very fast when $f$ increases from 30% to about 80%, but then it decreases again dramatically. Unfortunately, the optimal value of $f$ depends critically on the shell structure of the nucleus; it is rather high for magic nuclei (as superdeformed $^{152}_{66}\text{Dy}_{86}$) which have large shell gaps, and has to be lowered when there are several single-particle states near the Fermi surface. In practical calculations it is more efficient to use rather low values of $f$, and ensure convergence independently of the shell structure, than to risk a divergence. Usually, $f$=50% can be a safely recommended value.

The total CPU times required to solve the HF equations depend not only on the convergence rate governed by $f$, but also (of course) on the overall speed of the computer and (very importantly) on the performances of the specific compiler optimization procedures. This is illustrated in Fig. 2, where the relative times required to perform specific tasks are shown in percent of the total time. The total CPU times required to calculate the complete superdeformed band in $^{152}_{66}\text{Dy}_{86}$ range from one hour on the Cray C-90 supercomputer to several hours on fast workstations (SG Power Challenge L or IBM RS/6000), and to about 36 hours on a typical workstation (SUN-Hyper). However, these times result from averaging very different performances pertaining to specific tasks. Typically, three tasks take most of the time, namely, the calculation of the HO matrix elements of the mean fields, the diagonalization of the Routhians, and the calculation of the Coulomb field. On a vector computer, the first of these tasks takes relatively long time (40%) because it is composed of multiple short loops,[18] whereas the last one contributes very little (5%) due to rather long loops it requires. On superscalar computers these three tasks take almost the same times of about 20% each. The example of typical workstation shown in Fig. 2 illustrates the importance of the optimization options; here the compiler failed to optimize the subroutine calculating the mean fields and therefore this task takes now almost 30% of the CPU time.

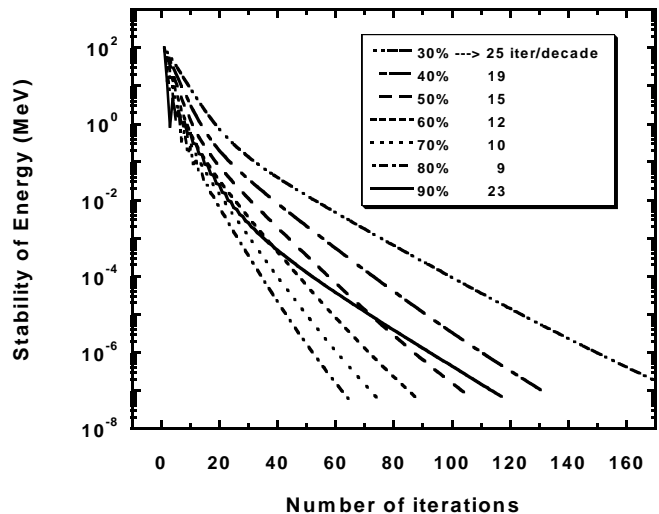

Figure 1: Stability energy as function of number of iterations for the calculations performed for the superdeformed state of $^{152}_{66}$Dy$_{86}$. Different curves correspond to different values of the fraction $f$ (see text), and lead to different numbers of iterations which are necessary to improve the stability by an order of magnitude.

## 4  Superdeformed bands in the $A\sim 60$ nuclei

Very recently there have been several reports of superdeformed rotational bands discovered in nuclei around $A\sim 60$.[23,24,25,26] Specific nuclei in this region have already been addressed from the point of view of the Nilsson-Strutinsky[27] and Relativistic Mean Field [28] descriptions. The single-particle structure of the light superdeformed nuclei is very simple and thus amenable to a systematic description in terms of a microscopic self-consistent mean-field approach. Using the code HFODD (v1.75),[29] hundreds of bands have recently been calculated in light Ni-Ga nuclei.[30] Based on these results one can find and analyze the generic features exhibited by rapidly rotating nuclei. In the present communication we report on quadrupole polarizabilities characterizing individual single-particle orbitals in this region.

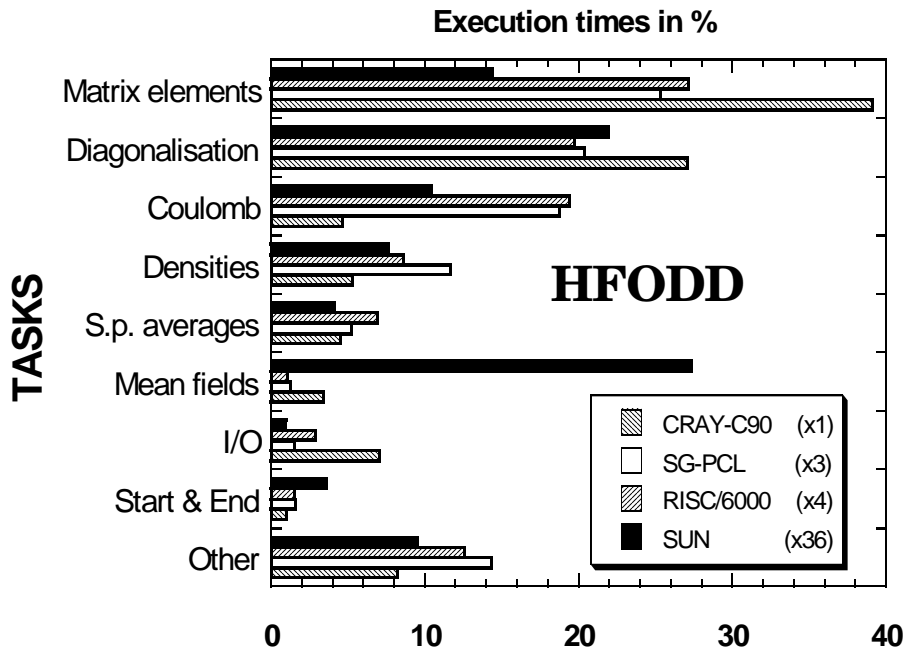

Figure 2: Percentages of the CPU time spent on performing separate tasks required by the HFODD code to solve the HF equations.

Following the method introduced in the $A\sim 150$ nuclei [31] we aim at a description of the proton quadrupole moments $Q_{20}$ in terms of contributions given by the particle (p) and hole (h) states defined with respect to the magic superdeformed nucleus $^{60}_{30}\text{Zn}_{30}$, i.e.,

$$Q_{20} = Q_{20}\left[^{60}_{30}\text{Zn}_{30}\right] + \sum_p \delta Q^p_{20} n_p + \sum_h \delta Q^h_{20} n_h \qquad (12)$$

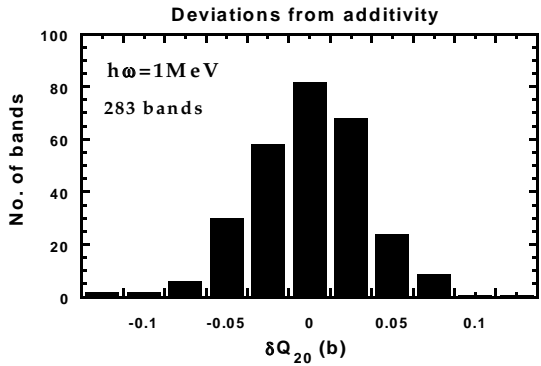

Figure 3: Deviations of calculated HF proton quadrupole moments in $A{\sim}60$ nuclei from the sum of individual single-particle contributions (12).

The core $^{60}_{30}$Zn$_{30}$ corresponds to all $n_p$=0 and all $n_h$=0, while the other superdeformed states in this region can be obtained by creating up to 6 particles with $n_p$=1 and/or up to 6 holes with $n_h$=1, both for protons and for neutrons.[30] Hence, the sum in Eq. (12) contains altogether 24 unknown coefficients $\delta Q_{20}$ which are fitted by the least-square method to reproduce all calculated proton quadrupole moments $Q_{20}$, separately at every rotational frequency $\hbar\omega$.

The quality of describing the proton quadrupole moments by the simple additive contributions is illustrated in Fig. 3 where we show deviations of the calculated HF values from the sum given in Eq. (12). One can see that for most of the bands the additivity hypothesis is correct up to 0.05 b. This constitutes a remarkable agreement with the values of $Q_{20}$ which are of the order of 2–3 b.

In Fig. 4 the values of individual contributions $\delta Q_{20}$ are shown as functions of the rotational frequency for the 24 orbitals considered. It seen that both proton and neutron orbitals significantly contribute to the proton quadrupole moment. This is so because the direct contributions of protons are of the similar order as the polarization contributions existing for both kinds of particles. The positive-parity orbitals [440]1/2 and [431]3/2, originating from the $1g_{9/2}$ intruder orbital, have a larger influence on the quadrupole moment than the negative-parity orbitals (note that the scale is twice expanded in the top panel of Fig. 4). Deformations gradually increase when consecutive $1g_{9/2}$ orbitals are occupied. The influence of proton positive-parity orbitals is about twice larger than that of their neutron counterparts. The negative-parity [303]7/2 orbitals, originating from the $1f_{7/2}$ extruder orbital, significantly contribute to the quadrupole moment too. In this case the neutron and proton contributions are almost equal, i.e., the direct proton contribution is very small. On the other hand, the proton contribution from the [312]5/2 orbital is much larger than that of the neutron counterpart; hence this orbital contributes mostly through the direct quadrupole moment.

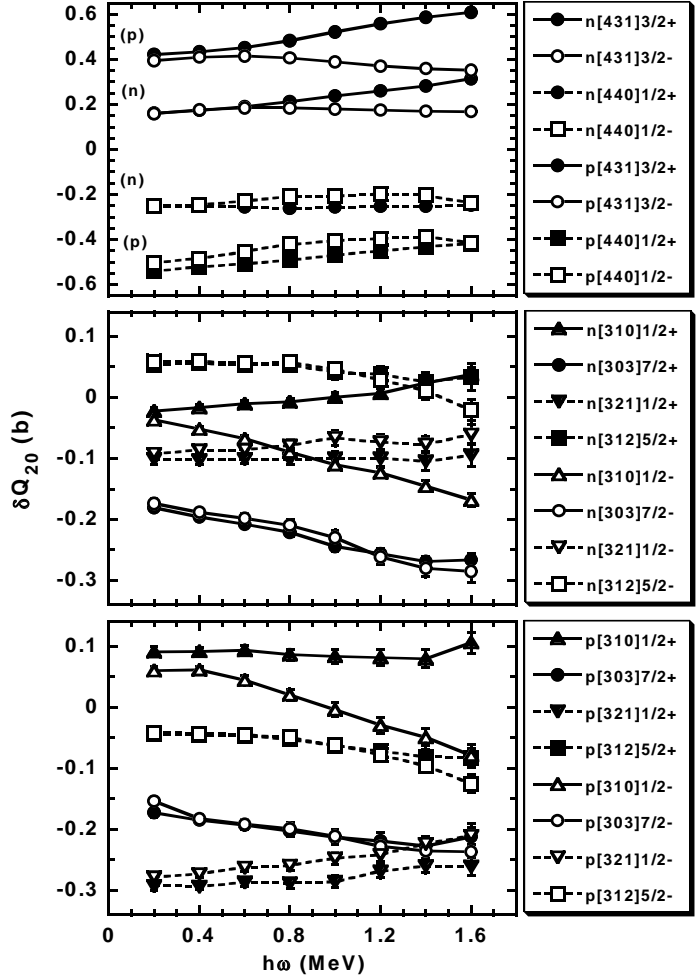

Figure 4: Contributions $\delta Q_{20}$ of individual single-particle orbitals to the proton quadrupole moments of nuclei in the $A{\sim}60$ region. The orbitals are denoted by the dominant Nilsson labels. Solid (dashed) lines correspond to particle (hole) states, while the closed and open symbols correspond to positive ($r=i$) and negative ($r=-i$) signatures, respectively. Error bars are determined by the least-square procedure.

Contributions coming from the signature-partner orbitals are almost equal, except from those related to the [431]3/2 and [310]1/2 orbitals where the positive-signature partners contribute significantly more. These two orbitals, and also [303]7/2, exhibit contributions slightly dependent on the angular frequency. However, these variations are relatively much smaller (do not exceed 0.2 b) than the variation of the core quadrupole moment $Q_{20}\left[^{60}_{30}\text{Zn}_{30}\right]$. Indeed, the latter decreases quite substantially with increasing $\hbar\omega$, from 3.28 b at $\hbar\omega$=0.2 MeV to 2.56 b at $\hbar\omega$=1.6 MeV. This is at variance with the results calculated for $^{152}_{66}\text{Dy}_{86}$, where the core quadrupole moment is fairly independent of spin.

## Acknowledgments

This work is supported in part by the Polish Committee for Scientific Research (KBN) and by the computational grants from the Interdisciplinary Center for Mathematical and Computational Modeling (ICM) of the Warsaw University and from the *Institut du Développement et de Ressources en Informatique Scientifique* (IDRIS) of CNRS, France (Project No. 960333).